**Kinetic asymmetry versus dissipation in the evolution of chemical systems as exemplified by single enzyme chemotaxis**

Niladri Sekhar Mandal,[a] Ayusman Sen,[a,b]* and R. Dean Astumian[c]*

Departments of Chemical Engineering[a] and Chemistry[b]

The Pennsylvania State University, University Park, Pennsylvania 16802 (USA)

Department of Physics and Astronomy,[c]

University of Maine, Orono, Maine 04469 (USA)

*Email: asen@psu.edu, astumian@maine.edu

**Abstract:**

Single enzyme chemotaxis is a phenomenon by which a non-equilibrium spatial distribution of an enzyme is created and maintained by concentration gradients of the substrate and product of the catalyzed reaction. These gradients can arise either naturally through metabolism, or experimentally, e.g., by flow of materials through several channels or by use of diffusion chambers with semipermeable membranes. Numerous hypotheses regarding the mechanism of this phenomenon have been proposed. Here we discuss a mechanism based solely on diffusion and chemical kinetics and show that kinetic asymmetry, a difference in the off rates for substrate and for product, and diffusion asymmetry, a difference in the diffusivities of the bound and free forms of the enzyme, are the sole determinates of the direction of chemotaxis. Exploration of these fundamental symmetries that govern nonequilibrium behavior helps to distinguish between possible mechanisms for the evolution of a chemical system from initial to the steady state, and whether the principle that determines the direction a system shifts when exposed to an external energy source is based on thermodynamics, or on kinetics, with the latter being supported by the results of the present paper. Our results show that while dissipation ineluctably accompanies non-equilibrium phenomena, including chemotaxis, systems do not evolve to maximize dissipation, but rather to attain greatest kinetic stability. Chemotactic response to the gradients formed by other enzymes provides a mechanism for forming loose associations known as metabolons. Significantly the direction of the effective force due to these gradients depends on the kinetic asymmetry of the enzyme, and so can be non-reciprocal, where one enzyme is attracted to another enzyme, but the other enzyme is repelled by the one, an important ingredient in the behavior of active matter.

**Introduction:**

Single enzymes and enzyme-coated microparticles have been shown to move directionally in gradients of small molecules (substrates)[1–5] resulting in a phenomenon which has been termed chemotaxis. This spontaneous movement in response to a gradient has potential applications in targeted drug delivery[6–8] the formation of dynamic assemblies[9,10], among many other possibilities. As enzymes translate along a substrate gradient, they inter-convert substrates and products as they dissipate chemical free-energy into the bulk environment. Chemical reactions taking place away from equilibrium and their associated thermodynamic irreversibility form the basis of living systems whose hallmarks are adaptation, self-assembly, and active motion. Therefore, an understanding of what controls enzyme chemotaxis can lead to important insights into how complex systems might evolve, what their stable states are, and what determines these stable states. Several hypotheses have been proposed in the literature to describe enzyme chemotaxis, including phoretic flows[11,12], thermodynamic drift forces[13], and cross-diffusion[3]. However, molecules in solution, including enzymes, move from place to place by Brownian motion, which when discussed in terms of the net motion of an ensemble of molecules is called diffusion. The movement of any single molecule in solution is characterized by a very low Reynolds number where the physical motion is a mechanical equilibrium process in which the viscous force is equal and opposite any inertial force acting on the molecule. Since every molecule is in mechanical equilibrium[14,15], how then is it possible to observe phenomenon such as single molecule chemotaxis that can occur only away from thermodynamic equilibrium? Answering this question, with a clear focus on the underlying sub-question of what determines the direction of chemotaxis, is the subject of this paper. The theoretical framework for our investigation is provided by trajectory thermodynamics[16–18], a theory based on the work of Onsager and Machlup[19] and later developed by Terrell Hill[20,21] for the description of enzyme catalytic cycles. In contrast to standard thermodynamic approaches which focus on the states of a system, trajectory thermodynamics focusses on the trajectories between the states, and on the relations between these trajectories provided by the principle of microscopic reversibility[22–24]. We show that the two essential components necessary for an enzyme to undergo chemotaxis in the presence of a gradient of substrate and product are diffusion asymmetry, where the bound and free states of the enzyme have different diffusion constants, and kinetic asymmetry[25], that is characterized by the difference in off rates of the product and the substrate. We contrast the role of kinetic and diffusive asymmetry with that of dissipation and demonstrate that the steady state enzyme distribution is not that which maximizes dissipation. Instead, it is kinetic asymmetry that determines the steady-state protein distribution.

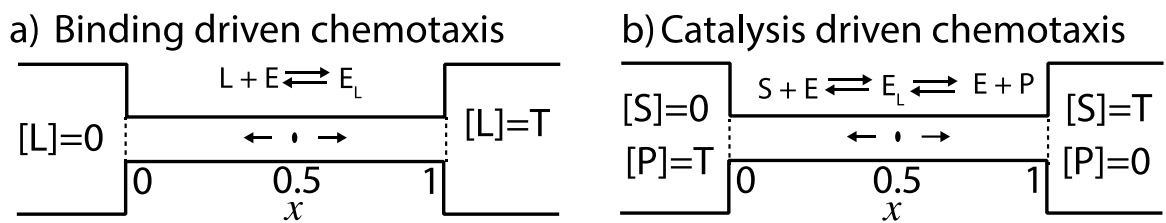

*Figure 1* Illustration of a single protein molecule E in a diffusion chamber with concentrations of ligands held fixed at the two ends of the chamber. The ligand molecules freely pass through a membrane on either end, represent by a dashed line, but the protein cannot pass through the membrane. In a), the protein has a simple binding reaction with the ligand, but there is no catalysis. The ligand concentration is fixed to be zero in the left reservoir, and we work in units such that the ligand concentration is T in the right reservoir. The diffusion constant, $k_D$ of the free (E) state of the protein may be different than the diffusion constant $k_{DL}$ of the bound form ($E_L$). In b) the protein is a catalyst – an enzyme – that facilitates the conversion of substrate S to product P. The concentration [S] is fixed at zero in the left reservoir and T in the right reservoir, and the concentration [P] is fixed at T in the left reservoir and zero in the right reservoir. The symmetric boundary conditions assure that once steady state has been established the sum of the concentrations at every point in the diffusion channel is a constant, $[S](x) + [P](x) = T$. We say that chemotaxis has occurred if, in the long-time average, the protein molecule spends more time to the left of the midpoint of the channel than on the right, or *vice versa*.

**Single enzyme chemotaxis:**

Consider the two cases shown in **Figure. 1**, where a single protein molecule is placed in a diffusion channel connected to reservoirs on the left and right, each with a membrane that is permeable to small ligand molecules, L, S, and P, but impermeable to the protein molecule. The situation shown in **Figure. 1a)** has been discussed recently[26] in the context of a protein molecule that can bind a ligand by the mechanism $L+E \underset{}{\overset{K_a^L}{\rightleftharpoons}} E_L$, where $K_a^L$ is the association constant. It was shown[12,26] that diffusion asymmetry where the diffusion constants of the bound and free forms of the protein differ from one another ($k_D \neq k_{DL}$) is a necessary and sufficient condition for biasing the overall Brownian motion of the protein molecule such that it spends more time on one side of the midpoint of the channel than on the other. In the context of an ensemble of protein molecules, the distribution is shifted to the left or right. If the diffusion constant of the bound form $E_L$ is less than that of the free form E the enzyme preferentially moves toward the reservoir with higher concentration of L, and if the diffusion constant of the bound form $E_L$ is greater than that of the free form E the enzyme preferentially moves toward the reservoir with lower concentration of L. The direction of chemotaxis is governed solely by the sign of the difference ($k_D - k_{DL}$) and does not depend on the association constant $K_a^L$.

In the present paper we examine chemotaxis due to catalysis of a chemical reaction as shown in **Figure. 1b)**. The protein in this case is an enzyme that catalyzes the reaction $S \rightleftharpoons P$ by the Michaelis-Menten mechanism $S+E \rightleftharpoons E_L \rightleftharpoons E+P$. Note that we write the bound form of the enzyme as $E_L$ rather than the more traditional $E_S$. Our convention is consistent with the observation by Hill[27] that substrate and product lose their individual identities (separate chemical potentials of bound substrate and bound product cannot be meaningfully defined) when bound to the enzyme. In order to separate equilibrium chemotaxis from the dissipative chemotaxis that is the focus here, we consider crossed equal gradients, where the concentration [S] is zero on the left and T on the right, and where the concentration [P] is T on the left and zero on the right. Once a steady state gradient of S and P has been established the total concentration $[S](x) + [P](x) = T$ is a constant and doesn't depend on the position $x$.

# Trajectory thermodynamics of single enzyme chemotaxis

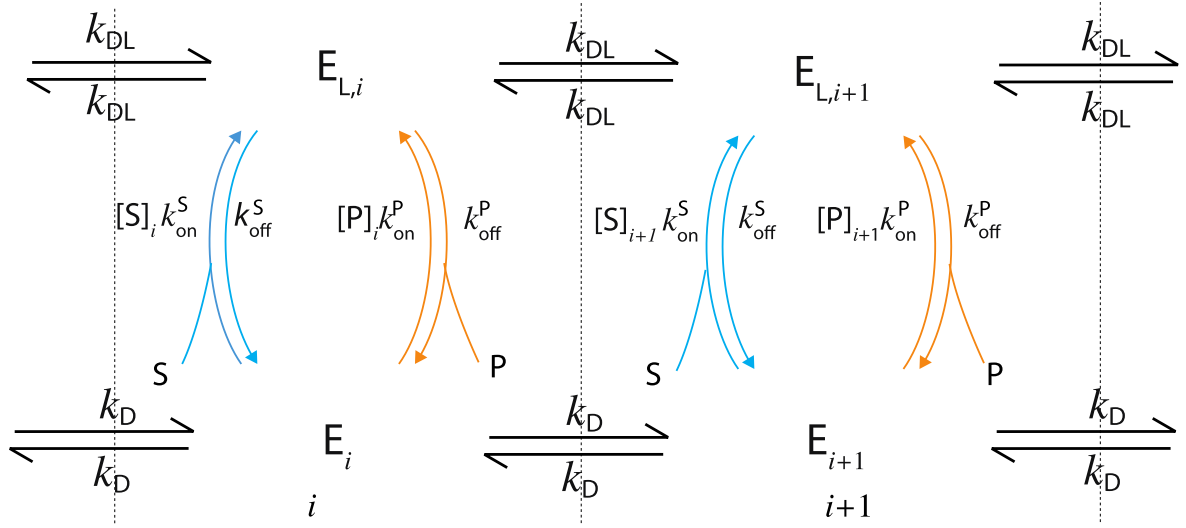

*Figure 2:* Enzyme reactions occurring in two adjacent chambers separated by an imaginary boundary. $E_i$ represents the unbound enzyme concentration in box $i$. $E_{L_i}$ represents the bound enzyme concentration in box $i$. $E_{i+1}$ represents the unbound enzyme concentration in box $i+1$. $E_{L_{i+1}}$ represents the bound enzyme concentration in box $i+1$. $[S]_i$ and $[P]_i$ represent substrate and product concentration in box i, respectively. $[S]_{i+1}$ and $[P]_{i+1}$ represent substrate and product concentration in box $i+1$, respectively. $k_D$ and $k_{DL}$ are the rates of diffusivity of the free and the bound enzyme species respectively. $k_{off}^S$, $k_{off}^P$ are the off-rates for the substrate and product respectively. $k_{on}^S$, $k_{on}^P$ are the bimolecular binding rate constants of the substrate and the product to the free enzyme respectively.

Consider the diffusion chamber shown in **Figure. 1b)** to be described in terms of N small compartments labelled $i = 0, \cdots, N-1$, shown in **Figure. 2**, where in each compartment the enzyme can carry out catalysis of a chemical reaction. The equilibrium constant for the catalyzed reaction is the ratio of the association constants for S, $K_a^S$, and for P, $K_a^P$, $K_{eq} = \frac{K_a^S}{K_a^P} = \frac{k_{on}^S}{k_{off}^S} \frac{k_{off}^P}{k_{on}^P}$, and where the enzyme can move between compartments by diffusion, where the diffusive steps are treated as unbiased thermally activated transitions denoted $k_D$ for diffusion of the free enzyme, and $k_{DL}$ for diffusive stepping of the bound enzyme.

Now we describe the possible paths between the bound state $E_{L,i}$ and $E_{L,i+1}$. In addition to the direct diffusive transition, $E_{L,i} \underset{k_{DL}}{\overset{k_{DL}}{\rightleftharpoons}} E_{L,i+1}$, which is modelled as an elementary kinetic step with equal forward and backward rate constants $k_{DL}$ there are four other paths (see Table 1) between the bound state $E_{L,i}$ and $E_{L,i+1}$.

***Table 1*** Forward and microscopic reverse trajectories between compartments $i$ and $i+1$.

| Trajectories $i \to i+1$ | Trajectories $i+1 \to i$ |
|---|---|
| $\mathcal{F}$: $E_{L,i} \xrightarrow{k_{\text{off}}^{P}} E_i \xrightarrow{k_D} E_{i+1} \xrightarrow{k_{\text{on}}^{S}[S]_{i+1}} E_{L,i+1}$ | $\mathcal{F}^\dagger$: $E_{L,i+1} \xrightarrow{k_{\text{off}}^{S}} E_{i+1} \xrightarrow{k_D} E_i \xrightarrow{k_{\text{on}}^{P}[P]_i} E_{L,i}$ |
| $\mathcal{B}$: $E_{L,i} \xrightarrow{k_{\text{off}}^{S}} E_i \xrightarrow{k_D} E_{i+1} \xrightarrow{k_{\text{on}}^{P}[P]_{i+1}} E_{L,i+1}$ | $\mathcal{B}^\dagger$: $E_{L,i+1} \xrightarrow{k_{\text{off}}^{P}} E_{i+1} \xrightarrow{k_D} E_i \xrightarrow{k_{\text{on}}^{S}[S]_i} E_{L,i}$ |
| $\mathcal{D}_S$: $E_{L,i} \xrightarrow{k_{\text{off}}^{S}} E_i \xrightarrow{k_D} E_{i+1} \xrightarrow{k_{\text{on}}^{S}[S]_{i+1}} E_{L,i+1}$ | $\mathcal{D}_S^\dagger$: $E_{L,i+1} \xrightarrow{k_{\text{off}}^{S}} E_{i+1} \xrightarrow{k_D} E_i \xrightarrow{k_{\text{on}}^{S}[S]_i} E_{L,i}$ |
| $\mathcal{D}_P$: $E_{L,i} \xrightarrow{k_{\text{off}}^{P}} E_i \xrightarrow{k_D} E_{i+1} \xrightarrow{k_{\text{on}}^{P}[P]_{i+1}} E_{L,i+1}$ | $\mathcal{D}_P^\dagger$: $E_{L,i+1} \xrightarrow{k_{\text{off}}^{P}} E_{i+1} \xrightarrow{k_D} E_i \xrightarrow{k_{\text{on}}^{P}[P]_i} E_{L,i}$ |
| $\mathcal{F}^*$: $E_i \xrightarrow{k_{\text{on}}^{S}[S]_i} E_{L,i} \xrightarrow{k_{DL}} E_{L,i+1} \xrightarrow{k_{\text{off}}^{P}} E_{i+1}$ | $\mathcal{F}^{*\dagger}$: $E_{i+1} \xrightarrow{k_{\text{on}}^{P}[P]_{i+1}} E_{L,i+1} \xrightarrow{k_{DL}} E_{L,i} \xrightarrow{k_{\text{off}}^{S}} E_i$ |
| $\mathcal{B}^*$: $E_i \xrightarrow{k_{\text{on}}^{P}[P]_i} E_{L,i} \xrightarrow{k_{DL}} E_{L,i+1} \xrightarrow{k_{\text{off}}^{S}} E_{i+1}$ | $\mathcal{B}^{*\dagger}$: $E_{i+1} \xrightarrow{k_{\text{on}}^{S}[S]_{i+1}} E_{L,i+1} \xrightarrow{k_{DL}} E_{L,i} \xrightarrow{k_{\text{off}}^{P}} E_i$ |
| $\mathcal{D}_S^*$: $E_i \xrightarrow{k_{\text{on}}^{S}[S]_i} E_{L,i} \xrightarrow{k_{DL}} E_{L,i+1} \xrightarrow{k_{\text{off}}^{S}} E_{i+1}$ | $\mathcal{D}_S^{*\dagger}$: $E_{i+1} \xrightarrow{k_{\text{on}}^{S}[S]_{i+1}} E_{L,i+1} \xrightarrow{k_{DL}} E_{L,i} \xrightarrow{k_{\text{off}}^{S}} E_i$ |
| $\mathcal{D}_P^*$: $E_i \xrightarrow{k_{\text{on}}^{P}[P]_i} E_{L,i} \xrightarrow{k_{DL}} E_{L,i+1} \xrightarrow{k_{\text{off}}^{P}} E_{i+1}$ | $\mathcal{D}_P^{*\dagger}$: $E_{i+1} \xrightarrow{k_{\text{on}}^{P}[P]_{i+1}} E_{L,i+1} \xrightarrow{k_{DL}} E_{L,i} \xrightarrow{k_{\text{off}}^{P}} E_i$ |

One, which we designate as $\mathcal{F}$, involves release of product, P, diffusion in the unbound state from $E_i$ to $E_{i+1}$ and binding of S to reach $E_{L,i+1}$. The microscopic reverse $\mathcal{F}^\dagger$ involves release of substrate from $E_{L,i+1}$, diffusion to $i$ in the unbound form, and binding P to reach $E_{L,i}$. There are three other paths in which the bound enzyme steps from compartment "$i$" to compartment "$i+1$". The probabilities $\pi_\mathcal{F}, \pi_\mathcal{B}, \pi_{\mathcal{D}_S}$, and $\pi_{\mathcal{D}_P}$, and $\pi_{\mathcal{F}^\dagger}, \pi_{\mathcal{B}^\dagger}, \pi_{\mathcal{D}_S^\dagger}, \pi_{\mathcal{D}_P^\dagger}$ for these trajectories and for their microscopic reverses are proportional to the products of the rate constants involved.

Similarly, in addition to the direct path between states $E_i$ and $E_{i+1}$ with forward and reverse rate constants $k_D$, there are four additional paths contributing to the transition $E_i \to E_{i+1}$ with probabilities $\pi_{\mathcal{F}^*}, \pi_{\mathcal{B}^*}, \pi_{\mathcal{D}_S^*}$, and $\pi_{\mathcal{D}_P^*}$, and $\pi_{\mathcal{F}^{*\dagger}}, \pi_{\mathcal{B}^{*\dagger}}, \pi_{\mathcal{D}_S^{*\dagger}}, \pi_{\mathcal{D}_P^{*\dagger}}$ for the probabilities of the microscopic reverses of those trajectories. The trajectories in Table I involve all states of the system and hence have identical normalization factors. This isn't true for the direct transitions, a fact that complicates the calculation of the general expression for the net transition probabilities $\pi_{+,i} = \sum[\pi(E_{L,i} \to E_{L,i+1}) + \pi(E_i \to E_{i+1})]$ and $\pi_{-,i+1} = \sum[\pi(E_{L,i+1} \to E_L) + \pi(E_{i+1} \to E_i)]$ and their ratio. However, we can obtain a bound by neglecting the sum $(k_D + k_{DL})$ comparison with the other terms in the net transition probabilities. In this case the expression simplifies to

$$\left|\ln\left(\frac{\pi_{+,i}}{\pi_{-,i+1}}\right)\right| \leq \left|\ln\left(\frac{\{k_{DL}[S]_i\mathcal{A}_i + k_D[S]_{i+1}\mathcal{A}_{i+1}\}}{\{k_D[S]_i\mathcal{A}_i + k_{DL}[S]_{i+1}\mathcal{A}_{i+1}\}}\right)\right| \quad (1)$$

Where we used the condition of microscopic reversibility for forward and microscopic reverse processes, $\frac{\pi_S}{\pi_{S^\dagger}} = e^{\mathcal{W}_S/RT}$ ($\mathcal{W}_S$ is the energy exchanged between the molecule and its environment, $\pm RT\ln\left(K_{eq}\frac{[S]_i}{[P]_i}\right)$ or 0 in the case studied here), and where

$$\mathcal{A}_i = \left[\frac{\frac{k_{off}^S}{k_{off}^P} + K_{eq}^{-1}\frac{[P]_i}{[S]_i}}{\frac{k_{off}^S}{k_{off}^P} + 1}\right] \quad (2)$$

is the kinetic asymmetry factor first discussed in the context of ATP driven molecular motors[28] but more recently recognized as being important for a wide variety of non-equilibrium phenomena including fuel driven assembly[29–31], directed motion,[32,33] and molecular adaptation[34].

The amount of energy dissipated in each conversion of S to P is $\Delta\mu_i = RT\ln\left(K_{eq}\frac{[S]_i}{[P]_i}\right)$. Clearly, dissipation plays a central role in catalysis driven chemotaxis since $\mathcal{A}_i = 1$ when $K_{eq}\frac{[S]_i}{[P]_i} = 1$. However, dissipation does not govern the direction of motion, which is controlled entirely by kinetics. If $k_D = k_{DL}$ the ratio of forward and backward steps is obviously one ($\frac{\pi_{+,i}}{\pi_{-,i+1}} = 1$), just as it is if $[S]_i\mathcal{A}_i = [S]_{i+1}\mathcal{A}_{i+1}$. If $[S]_i + [P]_i = T$ (as is the case in **Figure 1**) for all $i$, the latter identity holds if $k_{\text{off}}^{\text{S}} K_{\text{eq}} = k_{\text{off}}^{\text{P}}$. Broken diffusion symmetry and broken kinetic symmetry are both required for single enzyme chemotaxis, the direction of which is specified by the sign of $(k_{\text{D}} - k_{\text{DL}})(k_{\text{off}}^{\text{S}} K_{\text{eq}} - k_{\text{off}}^{\text{P}})$. If either term is zero, there is no chemotaxis. If the terms have the same sign, chemotaxis is in the direction of increasing substrate concentration and if they have opposite signs, chemotaxis is in the direction of decreasing substrate concentration. The process can be described as a one-dimensional random walk.

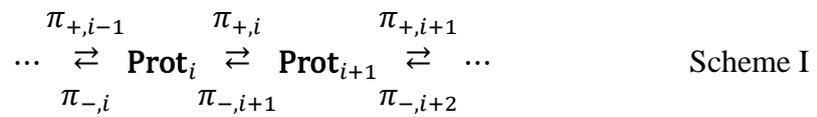

Scheme I

Where $\text{Prot}_i$ is the total enzyme at position $i$, with concentration $[\text{Prot}]_i = [\text{E}]_i + [\text{E}_\text{L}]_i$. We can understand how directed motion is engendered in the context of kinetic barrier[35,36] diagrams shown below (**Figure 3**),

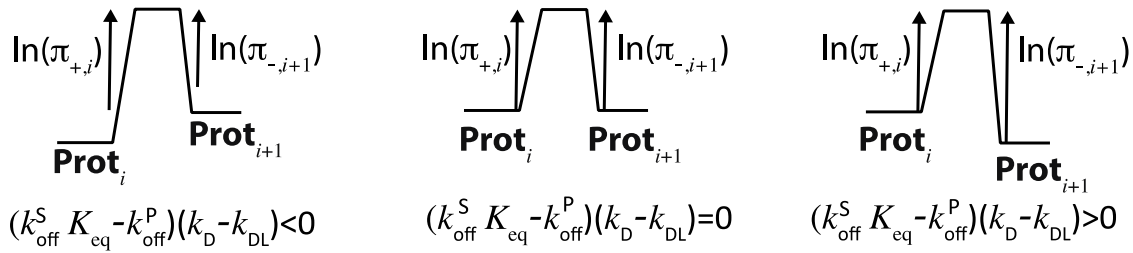

*Figure 3*: Kinetic barrier diagrams for three different conditions of the product of the diffusion and kinetic asymmetry. The net tilt depends on whether the product is positive, negative, or zero.

Depending on whether the product $(k_{\text{D}} - k_{\text{DL}})(k_{\text{off}}^{\text{S}} K_{\text{eq}} - k_{\text{off}}^{\text{P}})$ is less than, equal to, or greater than zero the overall landscape has a net tilt to the left, no tilt, or a net tilt to the right, respectively. The trajectory thermodynamic approach leading to Eq. (1) provides insight into the factors governing directionality and single enzyme chemotaxis, but to develop a quantitative understanding let us focus on a numerical solution of the kinetic equations (equivalently the reaction-diffusion equations) to determine how the enzyme concentration evolves under a varying substrate and product concentration with the constraint that $[\text{S}]_i +$

$[P]_i = T$ (we work in concentration units such that $T = 100$), and to calculate the steady state distribution.

The system evolves according to the differential equations for the free and bound enzyme in each compartment, respectively,

$$\frac{d[E]_i}{dt} = \overbrace{-(k_{on}^S[S]_i + k_{on}^P[P]_i)[E]_i + (k_{off}^S + k_{off}^P)[E_L]_i}^{\text{reaction}} + \overbrace{k_D([E]_{i-1} + [E]_{i+1} - 2[E]_i)}^{\text{diffusion}} \quad (3)$$

$$\frac{d[E_L]_i}{dt} = \underbrace{(k_{on}^S[S]_i + k_{on}^P[P]_i)[E]_i - (k_{off}^S + k_{off}^P)[E_L]_i} + \underbrace{k_{DL}([E_L]_{i-1} + [E_L]_{i+1} - 2[E_L]_i)} \quad (4)$$

Each equation comprises a reaction term describing the binding and release of substrate and product, and a diffusion term describing mass transport of protein from one box to another. The steady state levels can then be calculated by inverting the matrix of coefficients resulting from setting $\frac{d[E]_i}{dt} = \frac{d[E_L]_i}{dt} = 0$. Eqs. (3) and (4) can be used to explore the effect of arbitrary concentration gradients of S and P in the full parameter space. However, in the specific limits $(k_D - k_{DL}) = 0, (k_D - k_{DL}) \gg 0$, and $(k_D - k_{DL}) \ll 0$ we can develop analytic expressions for the steady state distribution of protein. We are searching for expressions for the total enzyme concentration in each cell, $[\text{Prot}]_i = [E]_i + [E_L]_i$ subject to the constraint that $\sum_{i=0}^{N}[\text{Prot}]_i V_i = \text{Prot}_{\text{Tot}}$ where $V_i$ is the volume of the $i^{\text{th}}$ cell and $\text{Prot}_{\text{Tot}}$ is the total amount of protein. For simplicity we take the volume of all cells to be identical. By adding Eqs. 3 and 4 we obtain the equation

$$\frac{d[\text{Prot}]_i}{dt} = k_D([E]_{i-1} + [E]_{i+1} - 2[E]_i) + k_{DL}([E_L]_{i-1} + [E_L]_{i+1} - 2[E_L]_i) \underbrace{= 0}_{\text{at steady state}} \quad (5)$$

where the reaction terms have cancelled in the sum. First, note that if $k_{DL} = k_D$ the steady-state solution requires $[E]_i + [E_L]_i = [\text{Prot}]_i = \text{Prot}_{\text{Tot}}/V$, a constant for all $i$, i.e., there is no chemotaxis irrespective of the other parameters. For $k_{DL} \neq k_D$, if $k_{DL} \to 0$, then $[E]_i$ has the same value for all $i$ ($[E]_i \equiv c_E$), an identity that can be used in Eq. (4) at steady state with $[S]_i + [P]_i = T$ to derive

$$\frac{[\text{Prot}]_i|_{k_{\text{DL}}\to 0}}{\text{Prot}_{\text{Tot}}} = c_{\text{E}}\left\{1 + K_a^P \frac{[(K_{\text{eq}}k_{\text{off}}^S - k_{\text{off}}^P)[S]_i + \text{T}k_{\text{off}}^P]}{k_{\text{off}}^S + k_{\text{off}}^P}\right\} = c_{\text{E}}(1 + K_a^S[S]_i \mathcal{A}_i) \quad (6)$$

If, on the other hand, $k_D \to 0$, then $[E_L]_i$ has the same value for all $i$ ($[E_L]_i \equiv c_{E_L}$) and we derive

$$\frac{[\text{Prot}]_i|_{k_D\to 0}}{\text{Prot}_{\text{Tot}}} = c_{E_L}\left\{1 + \frac{1}{K_a^P} \frac{k_{\text{off}}^S + k_{\text{off}}^P}{[(K_{\text{eq}}k_{\text{off}}^S - k_{\text{off}}^P)[S]_i + \text{T}k_{\text{off}}^P]}\right\} = c_{E_L}[1 + (K_a^S[S]_i \mathcal{A}_i)^{-1}] \quad (7)$$

Eq. 6 & 7 reiterate that, in addition to broken diffusion symmetry ($k_{\text{DL}} \neq k_{\text{D}}$), broken kinetic symmetry $(K_{\text{eq}}k_{\text{off}}^S - k_{\text{off}}^P) \neq 0$ is necessary for chemotaxis to occur. Plots of the protein concentration profile as a function of position are shown in **Figure. 4a)** and **4b)**. Both plots are monotonic – increasing to the right or left, with no maxima or minima.

*Effect of Inhibition or Activation on Enzyme Concentration.* Rate constants of several enzymes that are commonly used for chemotactic experiments are affected non-linearly by substrate or product concentration giving rise to inhibition or activation[37]. Incorporation of inhibition or activation by either substrate or product can lead to more complicated behaviors for the protein distribution versus position and even to a situation in which the distribution displays a maximum or minimum between the left and right semi-permeable membranes **(Figure 1)** as can be shown by plotting equations 6 and 7 with modified $k_{\text{off}}^{S,P}$ values.

To model inhibition by substrate, we set the catalytic rate (product off-rate) as a function of the substrate concentration, $k_{\text{off}}^P = \frac{\tilde{k}_{\text{off}}^P}{1+f([S])}$, where $\tilde{k}_{\text{off}}^P$ is the uninhibited off-rate of reaction at extremely small substrate concentrations. $f([S])$ is a function that models the dependency of the inhibition on the substrate concentration. If the inhibition is unaccompanied by a change in the substrate off-rate then according to the principle of microscopic reversibility the binding rates of the substrate or the product must change to keep the equilibrium constant of the overall reaction unchanged (since it is a property of the substrate and product free energies). We respect this symmetry by setting $k_{\text{on}}^P = \frac{\tilde{k}_{\text{on}}^P}{1+f([S])}$, where $\tilde{k}_{\text{on}}^P$ is the uninhibited binding rate. For this study, we use $f([S]) = [S] + [S]^2$ and then solve equations 3 and 4. We also take $k_S^{off} K_{eq} - \tilde{k}_{off}^P = 0$, i.e., the uninhibited kinetic asymmetry is set to zero.

As shown in **Figure. 4c)**, the kinetic asymmetry factor (which changes from cell to cell due to inhibition) acts as a potential function for the chemotactic behavior, with the maximum of the protein concentration coinciding with the minimum of $[S]_i \mathcal{A}_i$ (maximum of $([S]_i \mathcal{A}_i)^{-1}$ when $k_{DL} > k_D$ and maximum of $[S]_i \mathcal{A}_i$ when $k_D > k_{DL}$ (not shown), behaviors that are apparent from Eqs. (6) and (7). The protein diffuses such that its greatest probability occurs where the state of lower diffusion is maximally stabilized by the catalysis between S and P – i.e., the system seeks a state of greatest kinetic stability, not maximal dissipation.

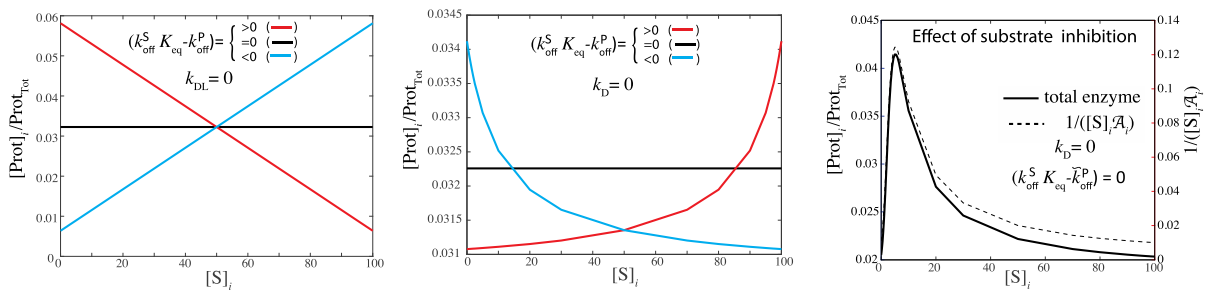

**Figure 4**: The variation of total enzyme when plotted against the substrate concentration for various values of diffusion and kinetic asymmetry. The total concentration of the substrate and product in each cell is set to be $[S]_i + [P]_i = T = 100$. **(a)** Variation of total enzyme concentration for $k_D = 1, k_{DL} = 0$. Kinetic asymmetry plays a key role in determining the direction of chemotaxis. **(b)** Variation of total enzyme concentration for $k_D = 0, k_{DL} = 1$. The direction of chemotaxis for the same kinetic asymmetry parameters is the opposite of what was seen for the faster free enzyme case. **(c)** Total enzyme distribution and $([S]_i \mathcal{A}_i)^{-1}$ plotted under substrate inhibition. The uninhibited kinetic asymmetry in the system is zero, corresponding to black lines in Figure 4(a) and 4(b). Clearly, the kinetic asymmetry factor $([S]_i \mathcal{A}_i)$ acts as a potential function for chemotactic behavior. The results shown in the figure were obtained using Eqs. 6 and 7 but were verified by solving the time dependent equations 3 and 4 to stationarity.

*Dissipation.* All non-equilibrium processes are accompanied by dissipation, and Prigogine designated those patterns originating from a flow of matter and/or energy as "dissipative structures"[38]. There are two contrasting proposals regarding the determining factor for non-equilibrium behavior. One proposal focuses on dissipation as a fundamental limit to precision of molecular machines and as a "driving force" that guides systems to adapt so as to maximize their dissipation[39,40]. The other proposal emphasizes the role of kinetics and in particular the requirement of kinetic asymmetry[25,31,41]. As a simple illustrative example, the diffusive chemotactic system outlined in the present paper highlights the correct prediction based on kinetic asymmetry versus the incorrect prediction based on dissipation. Let us focus on the case where $K_{eq} = 1$. The dissipation of an enzyme is zero at the center of the chamber where $[S]_i = [P]_i$, and increases toward the two reservoirs on the left and right. For $k_{off}^S = k_{off}^P$ the dissipation is maximized at positions where either of the substrate and the product concentrations are high (both ends of **Figure 1**) and is zero in the middle where $[S]_i = [P]_i$. In contrast, for $k_{DL} \ll k_D$, the protein occupancy increases or decreases linearly from left to right, depending on whether $(K_{eq}k_{off}^S - k_{off}^P)$ is greater than or less than zero. Thus, the hypothesis that dissipation plays a role in governing the steady state behavior of catalysis-driven chemotaxis is ruled out by the symmetry of the dissipation function in the system compared with the symmetry of the chemotactic behavior. While there is indeed dissipation due to the chemical catalysis, the direction of chemotaxis is governed solely by the product of the kinetic asymmetry and the diffusion asymmetry. When $(K_{eq}k_{off}^S - k_{off}^P)(k_{DL} - k_D)$ is positive, a single enzyme undergoes chemotaxis away from the reservoir with large substrate concentration and when $(K_{eq}k_{off}^S - k_{off}^P)(k_{DL} - k_D)$ is negative, a single enzyme undergoes chemotaxis toward the reservoir with large substrate concentration, and when $(K_{eq}k_{off}^S - k_{off}^P)(k_{DL} - k_D)$ is zero, there is no chemotaxis. We can summarize the behavior of an enzyme in a gradient of substrate and product by observing that the enzyme will spend most of its time in the region in which the less diffusive state (free or bound) is most favored by the kinetic asymmetry of the dissociation processes.

**Conclusion**

A perusal of the literature would suggest that enzymes and other molecular machines away from thermodynamic equilibrium – often captured under the sobriquet of "active matter" – can be characterized in terms of "violent kicks"[42], "judo throws"[43] and "leaping"[44].

The actual state of affairs is much gentler – the motion of these molecular systems in solution is best described as diffusion on an energy landscape[32,45,46] that is sculpted by evolution, by synthetic design, or, as in the present paper, by experimental setup. The behavior is consistent with the low Reynolds number regime in which viscous forces dominate inertial forces and where the motions of molecules are mechanically equilibrated processes. The mechanical equilibrium justifies the Markov description we have used in this paper to compare and contrast the relative roles of kinetic asymmetry and dissipation, which is relevant to the broader question of how simple matter becomes complex[47].

A thermodynamic hypothesis that systems adapt to maximize their dissipation, and that complex structures, in general, dissipate energy more rapidly than less complex structures has been propounded by England[39,48] and several others[49–52]. A second hypothesis is based on kinetics, and on how autocatalytic systems can give rise to a "dynamic kinetic stabilization" (DKS)[53] of structures. In the context of single enzyme chemotaxis, the DKS model predicts that the enzyme will move toward the reservoir near which the least diffusive enzyme species is favored by the kinetic asymmetry, a prediction borne out by the results shown in **Figure 4**. Thus, we conclude that kinetic and diffusion asymmetry governs enzyme chemotaxis by the simple mechanical equilibrium process of diffusion and have excluded the role of dissipation in determining the direction of chemotaxis. While diffusion is not commonly viewed as a chemical kinetic process, we treat diffusion as a thermally activated process which resembles a chemical reaction without a net change in free energy[54].

Enzyme chemotaxis can potentially find applications in several new technologies that require directed transport and targeted delivery, and most likely plays a role in the enhancement of catalytic activity in the presence of concentration gradients[55]. Further, chemotaxis of enzymes in the presence of concentration gradients of substrates and products provides a mechanism by which enzymes having common substrates/products can exert effective attractive or repulsive interactions. The treatment of chemotaxis in our paper extends to experimentally observed metabolons[2], loose associations of enzymes, a hallmark of enzyme cascades ubiquitous in living systems. Significantly, these interactions can be non-reciprocal[56] where a downstream enzyme whose substrate is the product of an upstream enzyme is attracted to or repelled by the upstream enzyme due to the effect concentration gradient of the product/substrate, but the upstream enzyme experiences no attraction or repulsion from the downstream enzyme. Notably, the interaction strength is not determined by the dissipation rate but depends instead on the diffusional and kinetic asymmetry of the downstream enzyme.

This role of kinetics has been discussed in the context of evolution of life by Pross and colleagues[57,58], but finds its intellectual roots in earlier work on kinetic asymmetry[25,28,41]. Kinetic asymmetry has been key in molecular designs allowing motions driven by energy from external modulation of the environment (e.g., externally enforced time dependent changes in the redox potential of the system) to be directional such that chemicals can be stored at high chemical potential on individual host molecules[59] or at surfaces[60]. The importance of kinetic asymmetry has also been discussed in driving self-assembly, molecular pumps and in several other contexts[30,61,62]. Each of these phenomena are stepping-stones in the evolutionary process. Unlike the thermodynamics of a specific chemical reaction, reaction kinetics can be influenced by the directed evolution of the catalyst[63]. We cannot help but think that single molecule chemotaxis and kinetic asymmetry played an outsized role in the evolution of life itself.

**References**

1.  Sengupta, S. *et al.* Enzyme molecules as nanomotors. *J Am Chem Soc* **135**, 1406–1414 (2013).

2.  Zhao, X. *et al.* Substrate-driven chemotactic assembly in an enzyme cascade. *Nature Chemistry* **10**, 311–317 (2018).

3.  Mohajerani, F., Zhao, X., Somasundar, A., Velegol, D. & Sen, A. A Theory of Enzyme Chemotaxis: From Experiments to Modeling. *Biochemistry* **57**, 6256–6263 (2018).

4.  Feng, M. & Gilson, M. K. Enhanced Diffusion and Chemotaxis of Enzymes. *Annual Review of Biophysics* **49**, 87–105 (2020).

5.  Zhang, Y. & Hess, H. Chemically-powered swimming and diffusion in the microscopic world. *Nature Reviews Chemistry* (2021).

6.  Joseph, A. *et al.* Chemotactic synthetic vesicles: Design and applications in blood-brain barrier crossing. *Science Advances* **3**, (2017).

7.  Wang, J. *et al.* Self-Propelled PLGA Micromotor with Chemotactic Response to Inflammation. *Advanced Healthcare Materials* **9**, 1–8 (2020).

8.  Somasundar, A. & Sen, A. Chemically Propelled Nano and Micromotors in the Body: Quo Vadis? *Small* **17**, 1–7 (2021).

9.  Patino, T., Arqué, X., Mestre, R., Palacios, L. & Sánchez, S. Fundamental Aspects of Enzyme-Powered Micro- and Nanoswimmers. *Accounts of Chemical Research* **51**, 2662–2671 (2018).

10. Ji, Y. *et al.* Macroscale Chemotaxis from a Swarm of Bacteria-Mimicking Nanoswimmers. *Angewandte Chemie - International Edition* **58**, 12200–12205 (2019).

11. Agudo-Canalejo, J. & Golestanian, R. Diffusion and steady state distributions of flexible chemotactic enzymes. *European Physical Journal: Special Topics* **229**, 2791–2806 (2020).

12. Agudo-Canalejo, J., Illien, P. & Golestanian, R. Phoresis and Enhanced Diffusion Compete in Enzyme Chemotaxis. *Nano Letters* **18**, 2711–2717 (2018).

13. Schurr, J. M., Fujimoto, B. S., Huynh, L. & Chiu, D. T. A theory of macromolecular chemotaxis. *Journal of Physical Chemistry B* **117**, 7626–7652 (2013).

14. Astumian, R. D. The unreasonable effectiveness of equilibrium theory for interpreting nonequilibrium experiments. *American Journal of Physics* **74**, 683–688 (2006).

15. Astumian, R. D. Design principles for Brownian molecular machines: How to swim in molasses and walk in a hurricane. *Physical Chemistry Chemical Physics* **9**, 5067–5083 (2007).

16. Astumian, R. D. Thermodynamics and kinetics of molecular motors. *Biophysical Journal* **98**, 2401–2409 (2010).

17. Astumian, R. D. Trajectory and Cycle-Based Thermodynamics and Kinetics of Molecular Machines: The Importance of Microscopic Reversibility. *Accounts of Chemical Research* **51**, 2653–2661 (2018).

18. Zuckerman, D. M. & Russo, J. D. A gentle introduction to the non-equilibrium physics of trajectories: Theory, algorithms, and biomolecular applications. *American Journal of Physics* **89**, 1048–1061 (2021).

19. Onsager, L. & MacHlup, S. Fluctuations and irreversible processes. *Physical Review* **91**, 1505–1512 (1953).

20. Hill, T. L. The linear Onsager coefficients for biochemical kinetic diagrams as equilibrium one-way cycle fluxes. *Nature* **299**, 84–86 (1982).

21. Hill, T. L. *Free Energy Transduction in Biology*. (Academic Press, 1977).

22. Lewis, G. N. A New Principle of Equilibrium. *Proceedings of the National Academy of Sciences* **11**, 179–183 (1925).

23. Astumian, R. D. Microscopic reversibility as the organizing principle of molecular machines. *Nature Nanotechnology* **7**, 684–688 (2012).

24. Blackmond, D. G. "If pigs could fly" Chemistry: A Tutorial on the Principle of Microscopic Reversibility. *Angewandte Chemie - International Edition* **48**, 2648–2654 (2009).

25. Astumian, R. D., Chock, P. B., Tsong, T. Y. & Westerhoff, H. V. Effects of oscillations and energy-driven fluctuations on the dynamics of enzyme catalysis and free-energy transduction. *Physical Review A* **39**, 6416–6435 (1989).

26. Mandal, N. S. & Sen, A. Relative Diffusivities of Bound and Unbound Protein Can Control Chemotactic Directionality. *Langmuir* 1–6 (2021) doi:10.1021/acs.langmuir.1c01360.

27. Hill, T. L. & Eisenberg, E. Can free energy transduction be localized at some crucial part of the enzymatic cycle? *Quarterly Reviews of Biophysics* **14**, 463–511 (1981).

28. Astumian, R. D. & Bier, M. Mechanochemical coupling of the motion of molecular motors to ATP hydrolysis. *Biophysical Journal* **70**, 637–653 (1996).

29. Ragazzon, G. & Prins, L. J. Energy consumption in chemical fuel-driven self-assembly. *Nature Nanotechnology* **13**, 882–889 (2018).

30. Das, K., Gabrielli, L. & Prins, L. J. Chemically Fueled Self-Assembly in Biology and Chemistry. *Angewandte Chemie - International Edition* **60**, 20120–20143 (2021).

31. Astumian, R. D. Kinetic asymmetry allows macromolecular catalysts to drive an information ratchet. *Nature Communications* **10**, 1–14 (2019).

32. Feng, Y. *et al.* Molecular Pumps and Motors. *J Am Chem Soc* **143**, 5569–5591 (2021).

33. Pezzato, C., Cheng, C., Stoddart, J. F. & Astumian, R. D. Mastering the non-equilibrium assembly and operation of molecular machines. *Chemical Society Reviews* **46**, 5491–5507 (2017).

34. Astumian, R. D. Stochastic pumping of non-equilibrium steady-states: How molecules adapt to a fluctuating environment. *Chemical Communications* **54**, 427–444 (2018).

35. Burbaum, J. J., Raines, R. T., Albery, W. J. & Knowles, J. R. Evolutionary Optimization of the Catalytic Effectiveness of an Enzyme. *Biochemistry* **28**, 9293–9305 (1989).

36. Astumian, R. D. Paradoxical games and a minimal model for a Brownian motor. *American Journal of Physics* **73**, 178–183 (2005).

37. Reed, M. C., Lieb, A. & Nijhout, H. F. The biological significance of substrate inhibition: A mechanism with diverse functions. *BioEssays* **32**, 422–429 (2010).

38. Nicolis, G. & Prigogine, I. *Self-Organization in Nonequilibrium Systems: From Dissipative Structures to Order Through Fluctuations*. (Wiley and Sons, 1977).

39. England, J. L. Dissipative adaptation in driven self-assembly. *Nature Nanotechnology* **10**, 919–923 (2015).

40. Gingrich, T. R., Horowitz, J. M., Perunov, N. & England, J. L. Dissipation Bounds All Steady-State Current Fluctuations. *Physical Review Letters* **116**, 1–5 (2016).

41. Astumian, R. D. & Robertson, B. Imposed Oscillations of Kinetic Barriers Can Cause an Enzyme To Drive a Chemical Reaction Away from Equilibrium. *J Am Chem Soc* **115**, 11063–11068 (1993).

42. Liphardt, J. Single molecules: Thermodynamic limits. *Nature Physics* vol. 8 638–639 (2012).

43. Vale, R. D. & Milligan, R. A. The way things move: Looking under the hood of molecular motor proteins. *Science (1979)* **288**, 88 (2000).

44. Jee, A. Y., Dutta, S., Cho, Y. K., Tlusty, T. & Granick, S. Enzyme leaps fuel antichemotaxis. *Proc Natl Acad Sci U S A* **115**, 14–18 (2018).

45. Astumian, R. D., Mukherjee, S. & Warshel, A. The Physics and Physical Chemistry of Molecular Machines. *ChemPhysChem* 1719–1741 (2016) doi:10.1002/cphc.201600184.

46. Astumian, R. D. *et al.* Non-equilibrium kinetics and trajectory thermodynamics of synthetic molecular pumps. *Materials Chemistry Frontiers* **4**, 1304–1314 (2020).

47. Lehn, J. M. Perspectives in chemistry - Steps towards complex matter. *Angewandte Chemie - International Edition* **52**, 2836–2850 (2013).

48. England, J. *Every Life on Fire: How Thermodynamics Explains the Origins of Living Things*. (Basic Books, 2020).

49. Valente, D., Brito, F. & Werlang, T. Quantum dissipative adaptation. *Communications Physics* **4**, 1–8 (2021).

50. te Brinke, E. *et al.* Dissipative adaptation in driven self-assembly leading to self-dividing fibrils. *Nature Nanotechnology* **13**, 849–855 (2018).

51. Tretiakov, K. V., Szleifer, I. & Grzybowski, B. A. The Rate of Energy Dissipation Determines Probabilities of Non-equilibrium Assemblies. *Angewandte Chemie* **125**, 10494–10498 (2013).

52. van der Helm, M. P., de Beun, T. & Eelkema, R. On the use of catalysis to bias reaction pathways in out-of-equilibrium systems. *Chemical Science* **12**, 4484–4493 (2021).

53. Pross, A. & Pascal, R. How and why kinetics, thermodynamics, and chemistry induce the logic of biological evolution. *Beilstein Journal of Organic Chemistry* **13**, 665–674 (2017).

54. Laidler, K. J., Glasstone, S. & Eyring, H. *The Theory of Rate Processes*. (McGraw Hill, 1941).

55. Chen, R., Neri, S. & Prins, L. J. Enhanced catalytic activity under non-equilibrium conditions. *Nature Nanotechnology* **15**, 868–874 (2020).

56. Bowick, M. J., Fakhri, N., Marchetti, M. C. & Ramaswamy, S. Symmetry, Thermodynamics and Topology in Active Matter. *Physical Review X* **12**, 10501 (2021).

57. Pross, A. *What is Life? How Chemistry Becomes Biology*. (Oxford Landmark Science, 2013). doi:10.4081/eb.2013.br1.

58. Danger, G., d'Hendecourt, L. L. S. & Pascal, R. On the conditions for mimicking natural selection in chemical systems. *Nature Reviews Chemistry* **4**, 102–109 (2020).

59. Qiu, Y. *et al.* A precise polyrotaxane synthesizer. *Science (1979)* **368**, 1247–1253 (2020).

60. Feng, L. *et al.* Active mechanisorption driven by pumping cassettes. *Science (1979)* **374**, 1215–1221 (2021).

61. Amano, S., Fielden, S. D. P. & Leigh, D. A. A catalysis-driven artificial molecular pump. *Nature* **594**, 529–534 (2021).

62. Amano, S., Borsley, S., Leigh, D. A. & Sun, Z. Chemical engines: driving systems away from equilibrium through catalyst reaction cycles. *Nature Nanotechnology* **16**, 1057–1067 (2021).

63. Arnold, F. H. Design by Directed Evolution. *Accounts of Chemical Research* **31**, 125–131 (1998).

**Code Availability**

MATLAB code used to calculate steady state distribution of chemotactic systems is available from the corresponding authors upon reasonable request.

**Acknowledgments**

NSM and AS gratefully acknowledge funding of the research by the Air Force Office of Scientific Research FA9550-20-1-0393.

**Author contribution**

The work was conceived by all three authors. RDA and NSM developed the theory and simulations. All authors contributed to the discussion of results and writing of the manuscript.

Corresponding Author E-mail: asen@psu.edu (AS), astumian@maine.edu (RDA)

**Competing interests**

All authors declare they have no competing interests.